# IT outsourcing decision factors in research and practice: a case study

**Mohammad Mehdi Rajaeian**
School of Management and Enterprise
University of Southern Queensland (USQ)
Toowoomba, Australia
Email: MohammadMehdi.Rajaeian@usq.edu.au

**Aileen Cater-Steel**
School of Management and Enterprise
University of Southern Queensland (USQ)
Toowoomba, Australia
Email: aileen.cater-steel@usq.edu.au

**Michael Lane**
School of Management and Enterprise
University of Southern Queensland (USQ)
Toowoomba, Australia
Email: Michael.Lane@usq.edu.au

## Abstract

The concurrent effect of various internal and external factors on IT Outsourcing (ITO) decision making has seldom been investigated in a single study. Furthermore, research on external factors is scarce and there is no comprehensive theory that can fully explain ITO decisions made in practice. This paper explains how key decision factors, both internal and external, influence ITO decision making in a large Australian University. We also tested the feasibility of a highly regarded descriptive model of ITO decisions as the basic foundation of an ITO decision theory. The model failed to fully explain ITO decisions in our case organisation. We draw researchers' attention to the need for more exploration of external factors as well as clarification of contingency factors that may explain inconsistencies between ITO decision theories and practice, and call for more research for 'practicable' ITO decision aids. Implications for practice are also discussed in the paper.

## 1   INTRODUCTION

While the origins of information systems outsourcing can be traced back to 1960s (Dibbern et al. 2004) it was Kodak's 1989 contract with IBM (Applegate and Montealegre 1991) that led to the widespread interest in outsourcing. IT outsourcing started as a mechanism to lower costs, has grown steadily and is now a widely accepted practice in strategic management of IT. Most ITO decisions are very complex (Brannemo 2006) because they involve many decision factors (Ang and Cummings 1997), both technological and business factors (Gulla and Gupta 2011), some of them with uncertain value (Zhang et al. 2012), very convoluted interrelationships among the factors (Liu and Quan 2013) and the IT process itself is complicated (Jain and Thietart 2013). The emergence of outsourcing models where multiple vendors are dealing with multiple clients also increases the complexity of ITO decisions.

Research into ITO decision making is still scarce (Tamm et al. 2014) and our study is motivated by researchers' calls for a closer look inside the "black-box" of decision making (Blaskovich and Mintchik 2011) to understand the complexity of ITO choices. We seek to answer the following research questions:

RQ1. *How structured and formal are ITO decision-making processes in practice?*

RQ2. *To what extent do current theories of ITO research (e.g. Lacity et al.'s 2010 model) explain the internal and external determinants of ITO decisions in practice?*

RQ3. *To what extent does the case organisation use academic ITO research to inform and support its ITO decisions?*

This research in progress paper provides an overview of ITO determinant factors (both internal and external factors) from the literature and describes how these factors affect ITO decision making in practice, based on the findings from an in-depth analysis of the ITO decision making in one large organisation. The conclusion section outlines implications for theory and recommendations for practitioners and well as describing the future phases of the current study.

## 2   LITERATURE REVIEW

The high level of complexity associated with ITO has led to the use of various theories from different disciplines to study ITO decisions. These include economic theories (e.g. Transaction Cost Economics, Agency Theory), social and organisational theories (e.g. Diffusion of Innovation, Social Exchange Theory), strategic theories (e.g. Resource-Based Theory, Game Theory) and other theories (e.g. Knowledge-Based Theory) (Lacity et al. 2010b). Researchers (e.g. Blaskovich and Mintchik 2011; Tiwana and Bush 2007) assert that any single theory cannot fully explain the complex practice of ITO. Recently researchers have expressed a need to develop an *indigenous* theory of IT outsourcing, since current theories are *borrowed* from other disciplines and none of them can individually address the complex field of ITO (Lacity et al. 2010b; Lacity et al. 2008; McIvor 2008).

A comprehensive literature review by Lacity et al. (2010b) identified 138 independent factors used by researchers in ITO research and categorised them into two sets of ITO decision factors and ITO outcome factors. Using factors that had been examined many times, Lacity et al. (2010b) suggested a descriptive model of ITO decisions (as shown in Figure 1) to serve at the basis of future indigenous ITO theory building.

Three independent factors (supplier competition, legal and political uncertainties, and ethnocentrism) were reported as environmental factors by Lacity et al. (2010b) In this study we also consider *mimetic* behaviour of organisations (taking successful peer organisations as role models to achieve similar results) as an external factor, since the source of influence is external to the organisation. The literature on external or environmental determinants of ITO decisions is scarce (Blaskovich and Mintchik 2011; Lacity et al. 2010b). Later we explain in detail the findings of both prior research and our case study about how these factors affect IT sourcing decisions in practice.

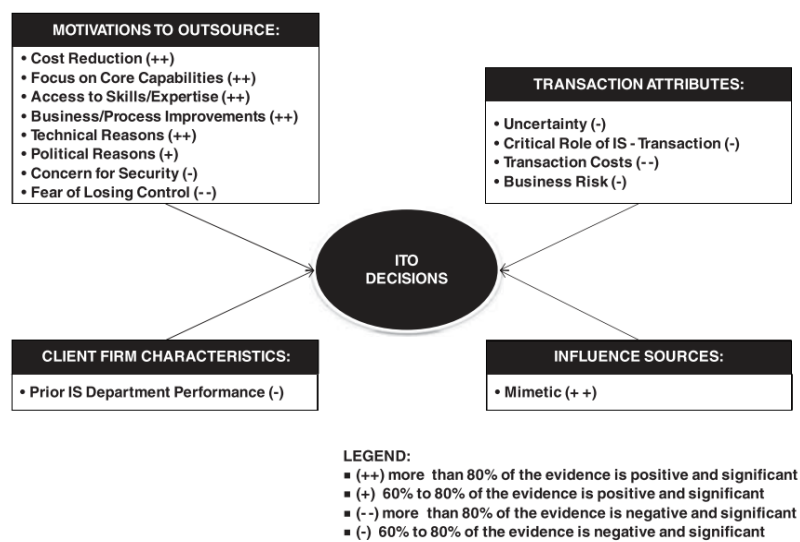

*Figure 1. Descriptive model of findings on ITO decisions (source: Lacity et al. (2010b))*

## 3　RESEARCH METHODOLOGY

We use a single organisation as a typical case (Saunders et al. 2011; Yin 2014) to demonstrate the synthesis of internal and external factors in the highly complex setting of ITO decision making. The purpose of our paper entails the inclusion of contextual information about the studied organisation to provide deep insights, thus a single case study method is preferred because in multiple-case research, there can be a high chance of losing valuable information by summarising multiple case studies (Peattie 2001).

We obtained approval (#H14REA103) from the USQ Human Research Ethics Committee and conducted semi-structured face-to-face interviews with four key people involved in ITO decision-making at USQ: the Strategic Procurement Administrator (Financial and Business Services), and three ICT executives: Deputy Vice-Chancellor (Academic Services) and Chief Information Officer; Director (Planning) of ICT Services; and Executive Director of ICT Services. Based on a review of the literature, we identified the factors that might affect ITO decisions and developed the questions for the interviews. Regarding internal determinant factors, the three IT managers were asked to fill in a questionnaire. We did not invite the Strategic Procurement Administrator to complete the questionnaire because many of the questions relied on expert IT knowledge and were not relevant to his position. We asked all participants about possible effects of each of the environmental factors identified. To enhance contextual understanding of the interview responses, we obtained and examined other data sources (website material, governmental documents, etc.). Two members of the research team conducted the interviews in May/June 2014. To maintain anonymity, the interviewees are coded and their comments are referenced to as (A) to (D) in this paper.

## 4　CASE STUDY

The University of Southern Queensland (USQ) is one of Australia's leading providers of on-campus and online (distance) education programs. USQ has about 1,654 staff members and more than 27,000 students. Because over 75 percent of students are studying online (USQ 2014) ICT infrastructure and services are absolutely vital to the university.

### 4.1　USQ ICT Governance Structure

The USQ ICT Governance Structure comprises the ICT Strategy Board supported by three committees: Information Standards Committee, ICT Services Committee and ICT Portfolio Committee. The ICT Strategy Board is the highest body responsible for defining University ICT strategy, ensuring alignment between the university strategy and all ICT activities within USQ. The ICT Strategy Board is an advice-giving committee that is involved in major IT decision making at USQ (C). The Division of ICT Services is positioned under the direct management of 'Deputy Vice-Chancellor Academic Services and Chief Information Officer' and consists of three departments: Planning; Infrastructure; and Client Services. Approximately 100 full time equivalent employees work in USQ's Division of ICT Services (A).

#### 4.1.1　History and trend of IT Outsourcing at USQ

The history of IT outsourcing at USQ goes back to the beginning of IT activities in USQ in the 1970s. Since then USQ has partnered with different IT service providers for various IT needs. In some areas such as systems development there has been a gradual move from traditional 'build everything in-house' to partnering with external IT providers. As a result of this approach, the capital budget has not grown as much as it would if USQ had developed all systems in-house. USQ is using cloud services and is actively considering wider use of cloud services because of their scalability, rapid delivery and potential to reduce cost (A).

#### 4.1.2　Volume and types of IT outsourcing at USQ

IT outsourcing at USQ involves various types of IT services including but not limited to: application management, student email, design and build of a data centre, support on server application and server deployment, security audit, desktop management, training packages, audio/visual installation (projectors, video conferencing, etc.) and database management (A). The current ICT budget at USQ is approximately $19 million per year, comprising $11.2 million of operational budget, $4 million of capital budget and around $3.8 million of recurrent expenditure that involves the cost of software licenses, maintenance, internet traffic, etc. The value of ITO in terms of budget is estimated at around $1 million per year (A). Cost of IT consultancies was approximately $0.4 million in 2014 (USQ 2014). The geographic location of USQ's ICT service providers includes a wide geographic pool of providers that extends from local (Toowoomba city) or State (Queensland) to other states of Australia and even some other countries (international) (A).

### 4.1.3 IT outsourcing decision making at USQ

There is no explicit formal policy and strategy with regards to ITO at USQ. Outsourcing decisions are made individually for each business case, as a part of the project management methodology (A; B; C; D). However, there are some sourcing polices documented in the Procurement and Purchasing Policy and Procedure, for instance: 'Before sourcing any goods or services externally, a Procurement Officer or Finance Officer will, in the first instance, investigate if the supply can be met from internal University sources' (USQ 2015).

A large number of different staff members can be involved in the process of decision making from the system sponsor (Functional Manager) to the project board, Legal Services, ICT Portfolio Committee, the ICT Strategy Board, CIO, Deputy Vice Chancellors, Vice Chancellor and the University Council depending on the size and specifications of the proposed project. The Portfolio Committee assesses proposed projects in light of university priorities and several criteria and recommends the priority of proposed projects to the ICT Strategy Board. The ICT Strategy Board gives advice to the CIO, and the CIO makes a recommendation to the Vice Chancellor for approval (B; C). Decisions about investments up to a certain value are made in the Divisions. Projects over that threshold require approval by the Vice Chancellor's Committee (C). External IT consultants may also be involved to support the ITO decision making (A; C). For instance, USQ has undergone under a voluntary audit by an external consultancy firm for readiness assessment for large-scale cloud-based arrangements (C).

A senior ICT executive responded to our question about the choice between seeking decision models and tools from academic literature versus from the practitioner's world as follows: 'Our inclination is to be looking at practice models … in practice we would probably take a rather pragmatic/practice orientation … I do think that procedure or rigour is important and asking the right questions is important … at the end of the day, these things are always value judgments, that is, we will ask ourselves whether we are comfortable with the potential of losing this application for a period of time, and what's the likelihood and what's the impact. So [for instance] for student email we made a decision [to outsource] … but staff email didn't [successfully] pass that [decision-making]' (C).

## 4.2 Internal Determinants of IT outsourcing at USQ

Table 1 shows a list of 20 possible *internal* determinants of IT outsourcing, derived from the literature (Lacity et al. 2010b; Lacity et al. 2011) and the opinions of three USQ ICT Managers about the impact level of these determinants on IT outsourcing decisions at USQ.

As Table 1 shows, there is a consensus among USQ ICT Managers on the impact level of three factors: 'focus on core capability' (high), 'political reasons' (low) and 'cost predictability' (high). However their opinions regarding 'cost reduction', 'commercial exploitation' and 'career development of IS employees' vary. With the exception of 'political reasons', 'commercial exploitation' and 'head count reduction', all factors are considered medium to very high impact determinants.

'Cost reduction' has remained the most important driver for a majority of client firms, from the earliest studies to more recent ones (Lacity et al. 2010b). However in this study 'cost reduction' received three different impact rankings. One of the interviewees believed that '… what we are doing more these days is partnering with ICT providers to give us access to cloud … so the upfront cost is reduced because we are not spending as much time developing something from scratch, … then we pay a maintenance fee … our capital budget is pretty much stable but our recurrent budget is increasing every year because we are implementing new systems. Student and staff numbers are increasing and all these things impact the license fee that we pay …' (C). Another ICT executive maintained that: '… cost reduction is almost never a key determinant … [with outsourcing] we can buy a set of skills at great expensive rates for short period of time rather than putting staff ourselves to do that. I don't think we are saving money doing that … with the exception of very commoditised services, the make-buy decisions we did on very big systems [shows] they are almost always exactly equal' (B).

Instead of 'headcount reduction', one manager reported 'headcount stabilisation' as a determinant (A). Another human resource issue in ITO decisions is the change in required skills, for instance by increasing the outsourcing level, the organisation needs new skills such as expertise in vendor management and less technical skills such as programming (C). Also levels of risk acceptance are different; 'there are universities that years ago adopted cloud' (C).

*Table 1. Determinants of ITO and their impact level on ITO at USQ*

| Motivation factor for Outsourcing | Impact level on IT outsourcing decisions | | | | |
|---|---|---|---|---|---|
| | Very High | High | Medium | Low | N/A |
| Cost reduction | | A | C | B | |
| Focus on core capability | | A, B & C | | | |
| Access to external expertise/skills | B | A & C | | | |
| Business/process performance Improvements | | A & C | B | | |
| Technical reasons (to gain access to leading-edge technology) | | A & B | C | | |
| Political reasons (to use an outsourcing decision to enhance their personal benefits) | | | | A, B & C | |
| Concern for security/intellectual property | C | A | B | | |
| Fear of losing control | | C | B | A | |
| Flexibility enablement | | A & C | B | | |
| Commercial exploitation (to partner with a supplier to commercially exploit existing client assets or form a new enterprise) | | | | B & C | A |
| Change catalyst | | B & C | | | A |
| Access to global markets (by outsourcing to suppliers in those markets) | | | B & C | A | |
| Scalability | B | A & C | | | |
| Rapid delivery | C | A & B | | | |
| Alignment of IS and business strategy | | A & C | B | | |
| Career development of IS employees | | B | C | A | |
| Cost predictability | | A, B & C | | | |
| Head count reduction | | A* | | B & C | |
| Innovation (to use outsourcing as an engine for innovation) | C | A & B | | | |
| * Head count stabilization rather than reduction  Each letter (A, B, C) represents one of interviewees. | | | | | |

### 4.2.1 IT outsourcing limitations at USQ

The IT Service Desk, staff email and the learning management system (LMS) are among those services that are provided internally at USQ (C). There are several reasons behind the decision to keep in-house some ICT infrastructure and services. Firstly, USQ has already significantly invested in ICT infrastructure. Because human resources and various workflows, processes and procedures have been successfully established within the organisation and are working well, managers see no rationale to outsource every IT infrastructure and service. The position of the organisation in its lifecycle also affects its outsourcing decisions. In other words, it could be possible for a newly established university to outsource the majority of its IT infrastructure and services especially by using cloud services (B; C), but not necessarily for a mature university. Human resources is a major factor here. Although organisations can employ in-placement (internal transfer) and out-placement (transferring to an external firm such as an outsourcing vendor) in addition to layoff strategy (Ranganathan and Outlay 2009), the degree of head count reduction is still limited due to the type or duration of some employees' contracts and legal limitations (D). Secondly, there is no viable solution available in the market at the moment for some required services (C). Thirdly, some of the systems (e.g. LMS) are considered strategic assets and maintaining them in-house gives USQ more flexibility and configurability (B). Moreover, USQ's experience has proved that an internal hosted LMS delivers better response time and more consistent uptime than some counterpart systems (e.g. Blackboard) hosted outside USQ in the region (B, C). Nevertheless, it would be possible that USQ outsources its LMS to the cloud in the future (C). Fourthly, USQ's ICT budget structure does not completely support outsourcing (C). Finally, certain data, e.g. student records and staff personnel records, should be stored locally or with a trusted host due to legal, privacy and security considerations (A; B; C; D).

## 4.3 Effects of External Factors on IT Outsourcing at USQ

This section presents various external factors identified from the literature that influence ITO decisions. The relevance of each factor to USQ's ITO decisions is discussed after the description of each category of factors.

### 4.3.1 Economic Environment

IT labour shortage, volatility in the IT job market and labour cost are major economic factors in the external environment that significantly impact ITO decisions.

IT labour shortage. One common global problem for many companies is the dearth of available professionals with key technical skills (Hätönen and Eriksson 2009). The shortage of skilled IT workers has been considered one of the key drivers of IT outsourcing (Lacity et al. 2010b). High demand and low supply of skilled IT human resources has been a recurring phenomenon in some countries. Consequently wages for IT staff and associated IT costs for organisations have increased. In such situations ITO and particularly offshoring provide a promising solution for many organisations (Chen et al. 2002).

Volatility in IS/IT job market. Some research (e.g. Slaughter and Ang 1996) showed that IS job positions with a volatile demand in the job market are more likely outsourced than those with a stable demand, and that jobs requiring relatively abundant skills are insourced more than jobs requiring scarce skills.

Labour costs. IT/IS projects are labour-intensive and are under vigorous time constraints. External pressure on an organisation forces it to consider outsourcing, particularly when events in the market, such as the increase in price competition, put pressure on organisations to reduce costs and to improve internal efficiency (Ruffo et al. 2007). It is claimed that the primary driver for offshore outsourcing and its many variations is cost reduction (Tjader et al. 2014).

Effect of economic factors on USQ's ITO. The local job market certainly has affected USQ's outsourcing decisions (B; C). USQ's main campus is located in regional Australia (Toowoomba) and even USQ's IT graduates tend to work in larger cities, thus the local supply of IT skills is limited. For instance, USQ has outsourced some tasks related to its computer network due to local unavailability of highly skilled Cisco network professionals (B). Three other USQ campuses give USQ another choice in IT sourcing decisions. For instance, Springfield campus is located near Brisbane and has better access to skilled IT human resources, and it is possible to employ people there, since the campuses are connected via network (A; C). Also, the variation in the IT job market depends on the Australian economy cycles, and affects USQ's IT sourcing decisions (A). Therefore sometimes sourcing locally is a challenge because the local business environment does not have the required skills and expertise (C).

### 4.3.2 Industry Environment

Supplier competition. Transaction Cost Theory (Williamson 1976) suggests that a high number of potential suppliers lowers switching costs in the event of terminating a contract, and thus positively affects ITO (Ang and Cummings 1997). In other words, the presence of competition reduces the vendor's bargaining power and prevents the vendor from locking in the customer (Lacity and Hirschheim 2003). On the other hand, a successful number of prior projects tends to increase the bargaining power of the vendor since this leads to a lock-in effect (Gopal et al. 2003). Outsourcing in a non-competitive market (or even relying on a single vendor in a competitive market) presents the organisation with a high degree of risk (Poston et al. 2009).

Mimetic influence. Organisations imitate their peers' initiatives based on the perception that the experience of peer organisations is an adoptable success story for them. At least five studies analysed

ITO from a "social system level" and found positive and significant effects on IT of mimetic behaviour on ITO decisions (Lacity et al. 2010b).

Industry maturity. One of our participants suggested industry maturity as a new determinant factor by reasoning that 'different sectors have different maturity ... higher education is moving fairly slowly, compared to other industries. I think some other industries went from in-sourcing to outsourcing in a year or two ... If we want to outsource all of ICT for the university, that would be a long slow painful process, because we have union engagement, government obligations, [and] compliance obligations ...' (B).

Effect of industry environment on USQ's ITO. ITO decisions and the experiences of other universities in Australia exert a significant influence on USQ's IT sourcing decisions. USQ ICT managers believe that 'the pressure is on and it is necessary to run the same applications or similar that other organisations in your business are running. So there is a competition factor ...' (D), 'the risk goes down when other universities have done the same thing successfully. We share information across the sector ... we are not competitive, we work together ... we actually even collaborate at a national level on sourcing contracts (B).

The participation of universities in the Council of Australian University Directors of Information Technology (CAUDIT) and also informal sharing of information and collaboration increases the mimetic effect (A; B). CAUDIT is an industry-wide group representing the IT Directors/CIOs of all universities in Australia and New Zealand. CAUDIT negotiates collective procurement agreements, provides professional development, undertakes projects and fosters collaboration through the sharing of ideas, experiences and best practice amongst its members. With 57 members and an annual IT expenditure of almost $2 billion, CAUDIT is able to speak authoritatively to government, industry and university bodies on all aspects of IT. CAUDIT also provides a strategic procurement service that provides significant value to members by leveraging collective spending power as well as building strategic partnerships. For instance CAUDIT has provided a discount of approximately 34 percent to members on Microsoft software purchases. In 2012 CAUDIT introduced a set of procurement guidelines to provide guidance and best practice when undertaking negotiations for collective procurement agreements with ICT vendors on a sector wide basis. In 2010, CAUDIT claims its procurement activities saved the Australian Higher Education sector more than $12 million (CAUDIT 2013).

### 4.3.3 Technological Environment

Technological uncertainty. A challenge for organisations is technological uncertainty and the risk of underestimating the costs of IS projects (McLellan et al. 1995). Through outsourcing, managers seek to reduce or mitigate these risks or/and transfer them to the supplier. The vendor organisation is expected to have superior skills and resources to manage these risks (Dibbern et al. 2004).

Emerging Technologies. Technology has continued to enable new sourcing models such as application service provision (ASP) and cloud computing. Cloud computing can be defined from the perspective of ITO as: "an enabler of IT outsourcing whereby access to IT resources such as software, hardware, and platforms are delivered over the Internet as a service and users are predominantly charged on a pay per-use basis" (Yigitbasioglu et al. 2013).

Effect of technological environment on USQ's ITO. USQ is currently using a wide range of cloud services such as: customer relationship management (CRM) solution, contract management system, plagiarism detection software, student email, payment gateway, electronic recruitment system, on-line training system, and electronic survey system. There is a trend towards expanding the use of cloud-based applications and services at USQ and more applications will be cloud-based in the near future. The increase in cloud sourcing has urged USQ to define a cloud computing policy (A; B).

### 4.3.4 Legal and Political Environment

Government regulations and restrictions. Examples of government regulations that affect IT sourcing decisions are: disallowing foreign firms from operating local telecommunication networks, or IT products/services import/export, and limiting internet-based information exchange activities (Tan et al. 1997). Some government regulations can be explicitly about IT sourcing decisions. For instance Queensland's 'Government Information Technology Contracting (GITC)' Framework (GITC 2014) contains standard contractual terms and conditions for use on the acquisition of ICT products and/or services and involves many aspects such as Intellectual Property & Moral Rights, Customer's Obligations, Contractor's Obligations, Privacy and Service Levels. Another example is the Queensland Information Privacy Act 2009 (IPA 2013) that contains a number of privacy principles that set out the rules for how personal information is to be collected, managed, used and disclosed by Queensland Government agencies (OIC 2012).

Legal and political uncertainties. Legal and political uncertainties in domestic markets may also drive sourcing decisions. Public perception and public's anti-offshoring views in countries such as the United States and the United Kingdom (Lacity et al. 2010a) might affect sourcing decisions within client locations. In addition, some governments require specific types of work to be done onshore. Unclear government rules and regulations such as vague attitudes toward trans-border data flows (Sobol and Apte 1995) also negatively affect ITO.

Intellectual property rights. Fundamental intellectual property rights in developing countries are relatively weak. Intellectual property protection such as trademarks and copyrights is a necessity for outsourcing activity (Chen et al. 2002). Therefore this affects the choice of country in offshore outsourcing.

Effect of Legal and Political environment on USQ's ITO. The USQ procurement process is based on Queensland's GITC Framework (A; D). USQ is also compliant with the Queensland IP Act (B). With the growing trend towards cloud services and applications at USQ, it is expected that privacy and security would be even greater concerns in the future.

### 4.3.5 Sustainability and ECO-factors

IT's crucial role in energy consumption and environmental related issues (Cai et al. 2013) has given rise to studies in areas such as Green computing, Green IT, Green ITO. In considering Green ITO, buyers and regulators expect providers to demonstrate strong capabilities in sustainability and there is a growing awareness of global standards, such as the Global Reporting Initiative (GRI), the Carbon Disclosure Project (CDP), ISO 26000 and the UN Global Compact (Babin and Nicholson 2009; Babin and Nicholson 2011).

Energy consumption and Greenhouse Gas (GHG) emissions. Most global ITO providers are major consumers of electrical power. The increasing power consumption by IT gives rise to environmental concerns related to ITO. Some researchers (e.g. Babin and Nicholson 2011) believe that global ITO does not solve the energy consumption problem, as it simply moves the problem from an in-house data centre to an outsourced facility.

Human environmental impact. The amount of travel (and related GHG emissions) and the amount of fresh water consumption by employees are two environmental impacts that can be measured. Given the global nature of outsourcing, air travel is the predominant mode and "travel may account for as much as half of the company's carbon emissions" (Miyoshi and Mason 2009). For global IT outsourcers, in areas of the world such as India where fresh water is not always plentiful, conservation of water may become an important sustainability issue since hiring thousands of personnel at global delivery centres creates significant water requirements (Babin and Nicholson 2011).

Effects of sustainability and eco-factors on USQ's ITO. USQ follows a sustainable procurement process as required under the Queensland Procurement Policy (2013) (D). Various initiatives such as recycling, waste disposal, and efficient energy use are covered in USQ's procurement policy to integrate sustainability into the procurement process. The environmental standards mandated by Australian government for the purchase of any ICT equipment and consumables set a minimum level of environmental performance and therefore pose a necessary condition for any supplier to respond to an ICT request for tender.

### 4.3.6 Sociocultural Environment

Language. Foreign language proficiency, skills of communication, and knowledge of foreign culture are valued highly in international outsourcing collaboration. Some research (Chen et al. 2002) suggests that culture and language obstacles are among the main reasons why the Chinese IT/IS market share is less in comparison to India in the global ITO market. Recent investment of Chinese government in English-language education (Lacity et al. 2010b) is one of its initiatives to reduce this barrier. This may have a considerable impact on the future IT offshoring market.

Cultural Distance. Research (Levina and Vaast 2008) shows that cultural differences are a more serious threat to effective outsourcing than organisational differences. Personal contact between local employees and the foreign employer may be culturally or politically sensitive (Tan et al. 1997). Also, employees from different cultures may struggle with accurate interpretation of implicit knowledge (Leonardi and Bailey 2008).

Ethnocentrism. Ethnocentrism may influence a client's preference to source, using domestic rather than nondomestic suppliers. An analysis of 263,000 bids from 31,000 providers from 70 countries on the website Rent-A-Coder.com (Gefen and Carmel 2008) showed that clients would prefer to hire a programmer from their own country. The only exception was that American clients preferred to outsource to their near neighbour Canada for programmers.

Effect of Sociocultural environment on USQ's ITO. Since the amount of offshore ITO is very low at USQ, and all of the offshore contracts are with English speaking countries, this study did not identify any socio-cultural factors impacting ITO decisions at USQ.

## 5  CONCLUSION AND IMPLICATIONS

ITO is now an inseparable part of IT management. As rightly mentioned by one of the managers (A), 'you can't do everything, so you have to outsource something, it is physically impossible for IT organisations to be masters of everything. So to some extent, you have to lose control or hand over control to a third party'. With the advent of cloud computing and global economic pressure, the trend towards outsourcing is increasing. ITO decisions are very complex organisational decisions involving numerous internal and external factors. This paper provided a comprehensive overview of these factors and how they impact IT sourcing decisions at USQ.

We categorised the external factors of IT outsourcing decision making into six different perspectives (economic, industry environment, technological, socio-cultural, legal and political, and sustainability and eco-factors) based on the existing literature and the key findings that emerged from our analysis of the case study of IT outsourcing decision making at USQ as shown in Figure 2. One of the key external factors (industry maturity) emerged from our case study.

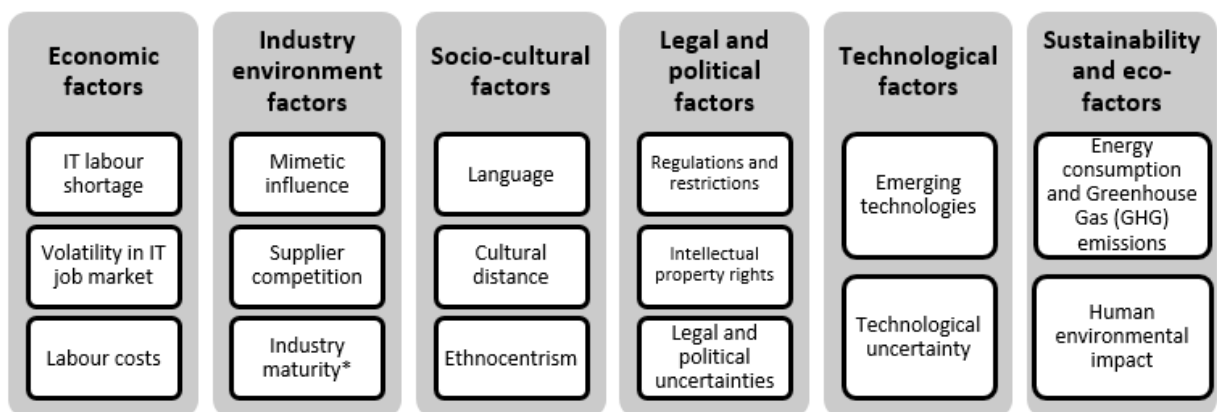

\* emerged from this study

*Figure 2. Six categories of environmental factors impacting on ITO (Source: developed from the literature for the purpose of this study)*

### 5.1  Implications for Theory and Practice

From a theoretical point of view our study contributes to the body of knowledge by highlighting the role of various internal and external factors in ITO decisions. In addition, the paper analyses the feasibility of current theoretical models of ITO decisions to evaluate their ability to explain and support ITO decisions in practice. Regarding the first research question, our study showed that there is no formal ITO strategy or systematic decision-making process specifically for ITO decisions at USQ.

To answer the second research question, as discussed in detail in the case study section, current theories of ITO research (e.g. Lacity et al.'s (2010) model) cannot thoroughly explain the ITO decisions in our case organisation. We found several inconsistencies, most noticeable of them was the lack of agreement on the 'cost reduction' factor. We also identified industry maturity level and the organisation's position in its lifecycle as pertinent factors that are seldom studied in ITO decision-making research. Mature organisations have certain challenges to adapt new sourcing strategies, particularly cloud computing, with regard to their previous investments both in IT human capital and IT infrastructure. Our study also highlighted the importance of the role of environmental factors in ITO decision making.

The third research question is also answered in the negative: no evidence was found of the use of any form of academic research-based decision knowledge in the organisation's decision process. This confirms prior research (e.g. Sven and Björn 2011; Westphal and Sohal 2013) that even though there are many proposed models for sourcing, organisations tend not to use these theoretical models and there is still a lack of decision support tools and frameworks to support ITO decision making in practice.

Based on the insights from ITO literature and our study of ITO in practice, a number of recommendations for practitioners are offered. First, collaboration between peer organisations in ITO can help practitioners to make complex ITO decisions and achieve enhanced outcomes. Past studies (Lane and Lum 2011; Whitley and Willcocks 2011) considered client/vendor 'collaboration' in a strategic sourcing context as proactive work to share the risks, and achieve high performance and mutually rewarding goals. Here we also draw practitioners' attention to another important type of collaboration: between clients. The collaborative practice of Australian Higher Education institutions can be a lesson for other organisations. Collaboration among outsourcing clients will increase both efficiency and effectiveness of ITO decisions by its 'cost reduction potential' (due to increased bargaining power); 'reduction in the bureaucracy' required to engage with vendors (e.g. common terms and conditions, a single price list and improved handling of procurement processes); and 'information and knowledge sharing' that leads to mitigate risk in ITO decisions and more effective decisions.

Second, we encourage organisations to develop a formal IT sourcing strategy integrated in the corporate ICT strategic plan. IT outsourcing is a strategic initiative and should be derived from a comprehensive strategy development process considering the role of current resources (both human capital and ICT infrastructure) and changes in the external environment. Budget model and human resource strategies need to be aligned with the organisation's IT sourcing strategies since with the expected increase in ITO and cloud sourcing, organisations require new skills and expertise such as vendor management and contract administration.

## 5.2  Limitations and Further Research

Our work should be interpreted in the light of its limitations. First, while the findings provide valuable insights into the evaluation of existing ITO theories, statistical generalisation is hardly possible due to use of a single case. For instance, the ITO decision-making behaviour in the case organisation cannot be generalised to the other Australian universities. Second, due to lack of prior studies that provide a comprehensive list of external factors, we used open questions in the interviews to elicit participants' views on the possible effects of the categories of factors (e.g. economic, technological, etc.). This introduced the possibility that participants may have failed to mention some factors, and also made it difficult to compare participants' views.

Further replications of this study will result in more generalisable results. We are currently conducting case studies of a four further organisations in different sectors to increase the reliability and validity of our findings regarding IT outsourcing decision making in organisations.

Also, future studies could focus on the exploration of external factors as well as clarification of contingency factors that may explain inconsistencies between ITO decision theories and practice. We also call for more research for 'practicable' ITO decision frameworks, models and tools to ensure organisations can fully benefit from ITO outsourcing.

## Acknowledgements

The authors wish to thank the four managers from University of Southern Queensland (USQ) for their participation and insights in this research.

## Copyright